\documentclass[aip,%
               preprint,%
               citeautoscript,%
               bibnotes,%
               amsfonts,%
               amssymb,%
               amsmath,%
               a4paper,%
               titlepage,%
               fleqn,%
               endfloats*%
               ]{revtex4-1}

\usepackage{graphicx}
\usepackage{xspace}
\usepackage{color}
\usepackage{caption,subcaption}

\definecolor{darkgreen}{rgb}{0.2,0.5,0.3}

\begin{document}

\newcommand{\hoch}[1]{$^{\text{#1}}$}
\newcommand{\tief}[1]{$_{\text{#1}}$}
\def\bA{{\bf A}\xspace}
\def\bB{{\bf B}\xspace}
\def\bC{{\bf C}\xspace}
\def\bD{{\bf D}\xspace}
\def\bE{{\bf E}\xspace}
\def\bF{{\bf F}\xspace}
\def\bG{{\bf G}\xspace}
\def\bH{{\bf H}\xspace}
\def\bI{{\bf I}\xspace}
\def\bS{{\bf S}\xspace}
\def\bT{{\bf T}\xspace}
\def\bL{{\bf L}\xspace}
\def\bR{{\bf R}\xspace}
\def\bU{{\bf U}\xspace}
\def\bV{{\bf V}\xspace}
\def\bX{{\bf X}\xspace}
\def\bY{{\bf Y}\xspace}
\def\bZ{{\bf Z}\xspace}
\def\bSig{\boldsymbol{\Sigma}\xspace}
\def\bDel{\boldsymbol{\Delta}\xspace}
\def\bNull{\boldsymbol{0}\xspace}
\def\bsig{\boldsymbol{\sigma}\xspace}
\def\btau{\boldsymbol{\tau}\xspace}
\def\bkap{\boldsymbol{\kappa}\xspace}
\def\bDbar{\boldsymbol{\overline{D}}\xspace}
\def\bDtil{\boldsymbol{\widetilde{D}}\xspace}
\def\br{{\bf r}\xspace}
\def\bb{{\bf b}\xspace}
\def\be{{\bf e}\xspace}
\def\bs{{\bf s}\xspace}
\def\bx{{\bf x}\xspace}
\def\bb{{\bf b}\xspace}
\def\bg{{\bf g}\xspace}
\def\ham{\hat{H}}
\def\hamp{\hat{\widetilde{H}}}
\def\w{\omega\xspace}
\def\gm{\mu\xspace}
\def\gn{\nu\xspace}
\def\gk{\kappa\xspace}
\def\gl{\lambda\xspace}
\def\bra#1{\langle #1|}
\def\ket#1{|#1\rangle}
\def\braket<#1|#2>{\langle #1| #2 \rangle}
\def\xc{\text{xc}}

\graphicspath{{./figs}}

\title{
A trust-region augmented Hessian implementation for restricted and 
 unrestricted Hartree--Fock and Kohn--Sham methods
}

\author{Benjamin Helmich-Paris}
\email{helmichparis@kofo.mpg.de}
\affiliation{Max-Planck-Institut f{\"u}r Kohlenforschung,
Kaiser-Wilhelm-Platz 1,
45470 M{\"u}lheim an der Ruhr,
Germany
}

\date{\today}

\begin{abstract}
We present a trust-region augmented Hessian implementation (TRAH-SCF) for restricted and 
 unrestricted Hartree--Fock and Kohn--Sham methods.
With TRAH-SCF convergence can always be achieved with tight 
 convergence thresholds, which requires just a modest number of iterations.
Our convergence benchmark study and our illustrative applications focus on
 open-shell molecules, also antiferromagnetically coupled systems, 
 for which it is notoriously complicated to converge
 the Roothaan--Hall self-consistent field (SCF) equations. 
We compare the number of TRAH iterations to reach convergence with those of
 Pulay's and Kolmar's (K) variant of the \emph{direct inversion of the iterative subspace} (DIIS)
 method and also analyze the obtained SCF solutions.
Often TRAH-SCF finds a symmetry-broken solution with a lower energy than DIIS and KDIIS. 
For unrestricted calculations, this is accompanied by a larger spin contamination, i.e.\ larger
 deviation from the desired spin-restricted $\langle S^2 \rangle$ expectation value.
However, there are also rare cases in which DIIS finds a solution with a lower energy than KDIIS and TRAH.
In rare cases, both TRAH-SCF and KDIIS may also converge to an excited-state determinant solution.
For those calculations with negative-gap TRAH solutions, standard DIIS always diverges.
If comparable, TRAH usually needs more iterations to converge than DIIS and KDIIS
 because for every new set of orbitals
 the level-shifted Newton-Raphson equations are solved approximately and iteratively by means 
 of an eigenvalue problem.
Nevertheless, the total runtime of TRAH-SCF is still competitive with the DIIS-based approaches
 even if extended basis sets are employed,
 which is illustrated for a large hemocyanin model complex.
\end{abstract}

\maketitle

\section{Introduction} \label{intro}

Most of today's quantum chemical calculations use either the Hartree--Fock (HF)
or Kohn--Sham (KS) density functional theory (DFT) ansatz to solve the molecular
Schr{\"o}dinger equation.
Typically, a solution is obtained with the self-consistent field (SCF) method  
 applied to the Roothaan--Hall (RH) equations\cite{Roothaan1951,*Hall1951}
that, until convergence is reached, repetitively diagonalizes the Fock matrix in the atomic-orbital (AO) basis
to receive a new set of molecular orbitals (MO).
The SCF approach often faces convergence issues and thus several tricks become necessary such as
mixing old and new density matrices\cite{Karlstroem1979,*Zerner1979} known as damping
or shifting the virtual-orbital energies by a constant or varying value\cite{Saunders1973,*Mitin1988}.
These two techniques are available and used by default in most quantum chemistry codes
 together with a extrapolation scheme known as the \textit{direct inversion of the iterative} subspace (DIIS).\cite{Pulay1980,*Pulay1982}

For most SCF calculations, the DIIS scheme together with damping and/or level shifts
 facilitates fast convergence within a few iterations.
However, following this route convergence cannot be guaranteed and often fails,
 in particular for open-shell molecules. 
Open-shell molecules often feature small energetic gaps between
 the highest occupied (HO) MO and the lowest unoccupied (LU) MO
 and multiple low-lying energy solutions exists.
Those obstacles lead to slow convergence rates or even divergence
 when following the standard SCF approach.

Alternatively, the HF and KS-DFT solution can be found with
 so-called second-order methods that employ the first and
 second derivative of the energy with respect to changes in the
 MO coefficients to minimize the energy by exploiting the variational principle.
A second-order quadratically convergent (QC) SCF method 
  was first proposed and implemented by Bacskay\cite{Bacskay1981,Bacskay1982}
 that can be still considered as state-of-the-art in many aspects:
(i) The orbital updates in every macro iteration were obtained
 from the eigenvalue equations of the gradient-augmented Hessian (AH)
 that contains the first and second energy derivative, respectively.
(ii) The eigenvalue equations were solved iteratively with the Davidson algorithm\cite{Davidson1975}
 and the accuracy of those micro iteration was coupled to the accuracy of the 
  current energy.
(iii) The linear transformations of the Hessian with trial vectors for the orbital
  update were implemented with a Fock matrix-based formulation
  that circumvents the costly $\mathcal{O}(N^5)$
  scaling transformations
  of the two-electron integrals from the AO to the MO basis
  with $N$ being a measure for the system size.

Though it was shown that the QC-SCF method was significantly more robust than
 the standard RH based SCF approach, there is still no guarantee that QC-SCF will always converge.
This issue was later resolved with a restricted step trust-region algorithm\cite{Fletcher1987} 
 for multi-configurational (MC) SCF methods by Jensen and J{\o}rgensen,\cite{Jensen1984,*Jensen1986,*Jensen1987}
 which at least in theory should always converge within a finite number of iteration.
Note that for MCSCF methods convergence is often even harder to obtain due to the coupling
 of the orbital rotation and configuration interaction parameters.
For that reason many other variants of second-order MCSCF optimizers 
 have been developed in the course of time.\cite{Yaffe1976,*Lengsfield1980,*Werner1980,*Werner1981,*Werner1985,*Shepard1982,*Sun2017,*Reynolds2018,*Kreplin2020}
However, that norm extended optimization algorithm of Jensen and J{\o}rgensen\cite{Jensen1984,*Jensen1986,*Jensen1987}
 is not directly applicable to single-determinant wavefunctions.
That is probably why J{\o}rgensen and his co-workers combined the AH approach of
 Bacskay\cite{Bacskay1981} with their previous restricted-step approach when working
 on a linearly scaling, completely AO-based trust-region augmented Hessian (TRAH) approach\cite{Salek2007,Hoeyvik2012a}
 for large closed-shell molecules.\cite{Hoest2008}

In the present work, we present a restricted-step trust-region implementation
 of the QC-SCF method that we call TRAH-SCF. 
Our primary focus is to accomplish robust convergence for both
 restricted closed-shell and unrestricted open-shell SCF calculations
 that are also still feasible for extended molecules with sufficiently large basis sets.
In Sec.\ \ref{sec:theory}, we will recapitulate the general TRAH method for
 finding minimum solutions of variational wavefunction models and also give some details
 for SCF approaches whenever necessary.
After presenting the computational detail in Sec.\ \ref{sec:compdet},
 we investigate the convergence behavior of closed-shell restricted
 and open-shell unrestricted HF and KS-DFT calculations in Sec.\ \ref{sec:scus-bench}
 with our TRAH-SCF implementation and compare it with the established DIIS and KDIIS methods.
For our convergence study, we use a small set of molecules with a notoriously complicated electronic structure
 and choose a single exchange-correlation (XC) functional for each of the following functional types: 
 local density approximation (LDA), generalized gradient approximation (GGA),
  hybrid with exact HF exchange, hybrid meta-GGA, and range-separated hybrid.
Also we investigate in Sec.\ \ref{sec:ru4co} the SCF convergence of unrestricted triplet and broken-symmetry
 singlet calculation for a small metal cluster that was studied previously\cite{Zeinalipour-Yazdi2008}
 also in the context of robust convergence.\cite{Hu2010,*Garza2012}
In Sec.\ \ref{sec:roussin} 
 we study the SCF convergence of an antiferromagnetically coupled 
 iron dimer system, i.e.\ Roussin's red dianion, with broken-symmetry unrestricted HF
 and DFT methods.
We also discuss spin contamination and symmetry breaking of the respective SCF solutions
 obtained with TRAH-SCF and the two DIIS implementations.
Then, we demonstrate the efficiency of our TRAH-SCF implementation for a hemocyanin model
 complex in Sec.\ \ref{sec:hemocy} and relate it to the conventional DIIS based implementations.
Finally, we conclude our work and give a perspective for future work in Sec.\ \ref{sec:concl}.

\section{Theory and Implementation} \label{sec:theory}

\subsection{ Trust-region augmented Hessian SCF method}

HF and KS-DFT are variational single-determinant electronic structure methods.
A solution of the electronic Schr{\"o}dinger equation for the electronic ground state $\ket{\tilde{0}}$
can be found by minimizing the energy with respect to the wavefunction parameters ${\bkap}$
\begin{align} \label{eq:schroe}
   E = \min_{\bkap} \, \bra{ \tilde{ 0 } } \hat{H} \ket{ \tilde{ 0 } } \\
\frac{ d E } {d \bkap} = \boldsymbol{ 0 }
 \text{.}
\end{align}
with $\hat{H}$ being the nonrelativistic molecular electronic Hamiltonian
 that may also include exchange-correlation potentials.
The final solution can then be described by 
 a unitary transformation from an arbitrary reference wavefunction $\ket{0}$
\begin{align} \label{eq:wfup}
 \ket{\tilde{0} } &= \exp( \hat{\kappa} ) \ket{0}
 \text{.}
\end{align}
In case of closed-shell restricted or open-shell unrestricted HF and KS-DFT methods, $\hat{\kappa}$
 represents rotations between inactive occupied and unoccupied virtual MOs.
So-called second-order methods expand the $E(\bkap)$ through second order in 
 the orbital-rotation parameters $\bkap$
and try to find an optimal solution for such a quadratic model 
\begin{align}
  Q(\bkap) = &E_0 + {\bkap}^T\, \bg + \frac{1}{2} {\bkap}^T\, \bH\, \bkap
\end{align}
to update the current wavefunction solution by invoking Eq.\ \eqref{eq:wfup}.
Minimizing the quadratic model for the SCF energy $Q(\bkap)$
requires the solution $\bkap$ of the Newton-Raphson linear equations
\begin{align} \label{eq:nr}
 & \bH\, \bkap = - \bg
\end{align}
for the orbital update.
Explicit equations for the electronic gradient $\bg$ and the linear transformations of
the Hessian $\bH$ are well known\cite{Gerratt1968} and recapitulated in the Appendix.

The energy expansion in terms of orbital rotations
 does formally not truncated after a finite order and the quadratic model is not necessarily a reasonable approximation,
 in particular, when being far from the final solution of Eq.\ \eqref{eq:schroe}.
This implies that finding optimal orbital rotations $\bkap$ by solving the NR equations \eqref{eq:nr} (micro iterations)
 and the subsequent orbital update Eq.\ \ref{eq:wfup} (macro iterations) has to be performed more than once.
The bigger obstacle is that in such situations the Hessian $\bH$ can be indefinite and/or 
 the NR solution gives a large step veering away from the final solution $\ket{\tilde{0}}$.
A remedy for such NR based optimization procedures is the introduction of a trust region, i.e.
 a sphere with radius $h$,
 in which the solution of the quadratic model is constrained to by
\begin{align} \label{eq:tradius}
 ||\bkap||_2 \le h
\text{.}
\end{align}
To find a solution subject to Eq.\ \eqref{eq:tradius},
 the minimum of the second-order Lagrangian
\begin{align}
 L(\bkap) &= Q(\bkap) - \frac{1}{2} \mu \left( \bkap^T \bkap - h^2 \right)
\end{align}
 has to be found, which results for the boundary condition of Eq.\ \eqref{eq:tradius}
  in the level-shifted NR equations
\begin{align} \label{eq:lsnr}
   \left( \bH - \mu \bI \right) \, \bkap = - \bg
 \text{.}
\end{align}
The shift $\mu$ is an additional unknown parameter that cannot be obtained
 by solving the linear equations \eqref{eq:lsnr}.
Instead, a value that is right below the lowest eigenvalue of the Hessian
 can be chosen to guarantee positive definiteness of the left-hand side of Eq.\ \eqref{eq:lsnr}.
An approximate determination of the shift $\mu$ prior to the solution of 
 the level-shifted NR equations \eqref{eq:lsnr} has been successfully attempted.\cite{Reynolds2018}
However, it is expected that the number of iterations for such a two-step approach
 are larger than for a one-step approach that determines all parameters $\mu$ and $\bkap$ simultaneously.
This can be achieved by diagonalizing the scaled augmented Hessian matrix
 \begin{align} \label{eq:trah}
  \begin{pmatrix}
             0  & \alpha \, {\bg}^T \\
\alpha \, \bg  &               \bH
  \end{pmatrix}
\,
  \begin{pmatrix}
  1 \\
\bkap(\alpha)
  \end{pmatrix}
=
 \mu \,
  \begin{pmatrix}
  1 \\
\bkap(\alpha)
  \end{pmatrix}
 \end{align}
for which level shift $\mu$  and orbital rotations $\bkap$ occur as
 lowest eigenvalue and the corresponding eigenvector, respectively.
The lower part of the eigenvalue equations \eqref{eq:trah}
\begin{align} \label{eq:lsnr-2}
 \left( \bH - \mu \bI \right) \, \frac{1}{\alpha} \bkap(\alpha) = - \bg
\end{align}
 resembles, apart from the scaling factor $1/\alpha$, the level-shifted NR equations \eqref{eq:lsnr}.
 The length of the update vector $\bkap$ can be chosen in such a way that 
 lies eihter within or on the surface of a trust-region sphere
\begin{align} \label{eq:constr}
 \frac{1}{\alpha^2} \, || \bkap(\alpha)||_2^2 \le h^2
 \text{.}
\end{align}
The constraint \eqref{eq:constr} can be imposed easily in an iterative Davidson algorithm\cite{Davidson1975}
for finding the lowest root of the AH in Eq.\ \eqref{eq:trah} (vide infra).

We note in passing that the particular form of the scaled AH eigenvalue equations is not unique.
In the mathematics literature, other eigenvalue equations for the the trust-region subproblem \eqref{eq:lsnr}
 were proposed and successfully implemented.\cite{Gander1989,*Sorensen1997,*Adachi2017}

\subsection{The iterative TRAH eigenvalue problem}

Concerning solving the TRAH eigenvalue equations, we follow
 Refs.\ \onlinecite{Salek2007,Hoeyvik2012a}
 and choose the following two initial trial vectors
\begin{align}
 \begin{array}{cc}
 \bb_0 = 
\begin{pmatrix}
 1 \\
 \boldsymbol{0}
\end{pmatrix} \text{,}
&
 \bb_1 = 
\begin{pmatrix}
 0 \\
 \bg / ||\bg||_2
\end{pmatrix}
\end{array}
\end{align}
This choice of trial vectors
separates the orbital update into two contributions ---
 one parallel to the gradient and the other orthogonal to it ---
 and leads to a  special structure of the reduced-space TRAH $\boldsymbol{\mathcal{A}}$
with
\begin{align}
 \boldsymbol{\mathcal{A}}(\alpha) 
 = 
\begin{pmatrix}
 0    &\alpha || \bg ||_2 & 0 & \cdots \\
 \alpha || \bg ||_2 & \mathcal{A}_{11} & \mathcal{A}_{12} & \cdots \\
 0  & \mathcal{A}_{21} & \mathcal{A}_{22} & \cdots \\
 \vdots & \vdots & \vdots &\ddots
\end{pmatrix}
\text{.}
\end{align}
To assist convergence to the lowest root, we add another start vector $\bb_2$
 that is zero for all elements except for the orbital rotation
 that matches the smallest virtual-occupied orbital energy difference. 
The latter is a reasonable approximation to the diagonal Hessian\cite{Chaban1997}
 and also employed as a preconditioner in the Davidson algorithm.
Trial vectors of all subsequent Davidson iterations are obtained from
 the residual of Eq.\ \eqref{eq:lsnr-2} after preconditioning and orthonormalization.
In our implementation, the Davidson algorithm is terminated if 
 the residual norm is below the norm of the current electronic gradient times
 a constant scaling factor $\gamma\, ||\bg||_2$ (see Tab.\ \ref{tab:para}).
By coupling the accuracy of the wavefunction update vector to that of the electronic
 gradient of the current solution, we avoid a wasteful accuracy of the orbital update vectors
 accompanied by an unnecessary large number of micro iterations.
Actually, because the electronic gradient of the current solution enters the
 Davidson procedure as a start vector the number of micro iterations in the beginning and
 end of the TRAH optimization is fairly similar in particular for the simple cases.

To keep the current orbital update vector within the trust region $h$,
 in every micro iteration the diagonalization of $\boldsymbol{\mathcal{A}}(\alpha)$
 is repeated until the constraint Eq.\ \eqref{eq:constr} is fulfilled.\cite{Salek2007,Hoeyvik2012a}
Here we only allow $\alpha$ values in a given interval $[\alpha_{\text{min}}, \alpha_{\text{max}}]$
  (see Tab.\ \ref{tab:para}) and perform a bisection search for $\alpha$ that gives the smallest
  deviation $h^2 - \frac{1}{\alpha^2} ||\kappa(\alpha)^{\text{red}}||_2^2$ with 
 $\kappa(\alpha)^{\text{red}}$ being the corresponding eigenvector of  $\boldsymbol{\mathcal{A}}(\alpha)$.
The equality condition $\frac{1}{\alpha^2} ||\kappa(\alpha)^{\text{red}}||_2^2 = h^2$
 giving the solution on the trust-region sphere
 usually occurs only for the first few macro iterations. 
Thereafter, $\alpha$ is located at the lower boundary $\alpha_{\text{min}}$
 and the step $\bkap(\alpha)$ is within the trust-region sphere, i.e.\ $\frac{1}{\alpha^2} ||\kappa(\alpha)^{\text{red}}||_2^2 < h^2$.

When the gradient norm is below a threshold (see Tab.\ \ref{tab:para}),
 our current solution is close to convergence and we switch to
 an iterative solution of the NR equations \eqref{eq:nr}
 without a level shift.
We use for this purpose a Davidson-type method customized for systems of linear equations.
Since the Hessian is positive definite when being close to convergence,
 we have also employed the often advocated preconditioned conjugate gradient (PCG) method.\cite{Hestenes1952}
But PCG did not offer any advantage for the shift-free NR equations because
 it turned out to be less robust for critical cases and did not show a faster convergence
 than the Davidson-type method.

\subsection{Orbital update}

%
%
Once the TRAH eigenvalue equations are solved approximately,
 the orbital rotation vector $\bkap$ is to compute the MOs $\mathbf{C}$ of the next macro iteration $k+1$,
  \begin{align}
  &\mathbf{C}^{k+1} = \mathbf{C}^{k} \, {\exp}( - \mathbf{K} ) \\[0.25em]
  &\mathbf{K} = 
\begin{pmatrix}
 \boldsymbol{0} & - \bkap^T \\
\bkap & \boldsymbol{0}
\end{pmatrix}
\text{.}
  \end{align}
By default, the matrix exponential
 is evaluated recursively by Taylor expansion.
 It is truncated once the norm of next expansion order is below $10^{-14}$.
To reduce the operations and hence the numerical error, we make use
 of the scaling-and-squaring approach\cite{Moler1978} with a fixed order ($s=3$) and 
 the exponent scale factor $1/2^s$.
Before the orbital update, a Cholesky orthogonalization of ${\exp}( - \mathbf{K} )$
 is performed to preserve orthonormality of the updated MOs.
Finally, the inactive occupied and virtual are transformed in such a way that
 MO Fock matrix becomes diagonal in their respective occupied-occupied and virtual-virtual subblocks.
This MO canonicalization usually improves the convergence rate of Davidson micro iterations.

\subsection{Step control}

In course of the TRAH optimization the trust radius should be
 updated to facilitate fast and robust convergence.
For this purpose, we employ Fletcher's algorithm\cite{Fletcher1987} that has been also chosen
 for many other restricted-step second-order implementations.\cite{Jensen1984,Jensen1996,Salek2007,*Hoeyvik2012a,Lipparini2016}
Based on the ratio 
\begin{align}
 r &= \frac{ E_{\text{actu}} }{ E_{\text{pred}} } \text{,} \\
 E_{\text{actu}} &= E^{k} - E^{k-1} \text{,} \\
 E_{\text{pred}} &= Q(\bkap) - E^{k-1} = \bg^T \bkap + \frac{1}{2} \bkap^T \bH \bkap \\
 &= \frac{1}{2} \left(  \bg^T \bkap + \frac{ \mu }{\alpha^2} \, ||\kappa(\alpha)^{\text{red}}||_2^2 \right) \text{,}
\end{align}
of the actual energy $E_{\text{actu}}$ and the predicted energy by the quadratic model $E_{\text{pred}}$,
the trust radius is potentially adjusted and the wavefunction update accepted or rejected:
\begin{itemize}
\item If $r < 0$, either the predicted or the actual energy rises 
 while the other falls. The quadratic model is not applicable within the
 given trust region and the new trust radius is decreased by $h^{k+1} =  0.7\,h^k$.
 Moreover, the recently updated orbitals are rejected and the micro iterations are
 repeated from the previous set of orbitals.
\item
If $0 \le r \le 0.25$, the recent update step was too long. Hence,
  the step is accepted but the new trust radius is decreased by $h^{k+1} = 0.7\,h^k$.
\item
If $0.25 < r \le 0.75$, the step is accepted and the trust radius is left unchanged.
\item
If $r > 0.75$, the step is accepted and the new trust radius is increased $h^{k+1} = 1.2\,h^k$.
 \end{itemize}

\section{Computational Details} \label{sec:compdet}

All calculations were performed with a development version of the ORCA
 quantum chemistry program package.\cite{Neese2012,*Neese2018,*Neese2020}
The new TRAH-SCF implementation will be publicly available in the upcoming release ORCA 5.0.
For our convergence studies, we compared our new TRAH-SCF implementation with 
 Pulay's original DIIS method\cite{Pulay1980,*Pulay1982} and Kolmar's variant\cite{Kollmar1997} (KDIIS).
For all our calculations, we only used default settings for a fair comparison
 that are documented for DIIS and KDIIS in the ORCA manual.
The default TRAH settings are compiled in Tab.\ \ref{tab:para}.
If not noted otherwise, all calculation started from MOs obtained from diagonalizing a Kohn-Sham matrix
  build from atomic-density contributions to the Coulomb matrix and exchange-correlation potential (\texttt{PModel}).

For the SCF convergence benchmark study with the Scuseria test set\cite{Daniels2000,Kudin2002} in Sec.\ \ref{sec:scus-bench},
 we used HF and the (i) Slater exchange\cite{Slater1951} with VWN5 correlation\cite{Vosko1980} (LDA); (ii)
 PW91 XC\cite{Perdew1992} (GGA); (iii) B3LYP XC;\cite{Becke1988,*Lee1988,*Becke1993}
 (iv) TPSSh XC;\cite{Tao2003} and (v) CAM-B3LYP XC functionals.\cite{Yanai2004}
The def2-TZVPP basis set was employed for the calculations on CrC, Cr\tief{2}, and NiC.
For UF\tief{4} and UO\tief{2}(OH)\tief{4} the scalar-relativistic second-order
 Douglas--Kroll--Hess (DKH2) Hamiltonian\cite{Douglas1974,*Jansen1989}
 and the DKH-SARC-TZVPP basis set\cite{Pantazis2011} were chosen for uranium.
 A customized version of the def2-TZVPP basis set\cite{Weigend2005} was used for the light elements
 with special contraction coefficients for the DKH2 Hamiltonian.\cite{PantazisDKH}
Also a tighter grid (\texttt{Grid7}) for the numerical integration of the XC functional
 was employed for all calculations of uranium-containing molecules.
All other KS-DFT calculations used the default grid.
The geometry of the five test molecules can be deduced from the parameter list in Ref.\ \onlinecite{Daniels2000}
 and from the UO\tief{2}(OH)\tief{4} structure provided as Supplementary Material.

For the convergence study on the Ru\tief{4}CO molecules,
 the B3LYP XC functional\cite{Becke1988,Lee1988,*Becke1993} was used together with the def2-TZVPP basis set.\cite{Weigend2005}
The structure was taken from a online repository stated in Ref.\ \onlinecite{Hu2010}.
The unrestricted broken-symmetry singlet calculations started from a unrestricted triplet calculation
 with a subsequent localization of quasi-restricted high-spin orbitals\cite{Neese2006b} (\texttt{BrokenSym}).

Calculations on the Roussin's red salt dianion ([Fe\tief{2}S\tief{2}(NO)\tief{4}]\hoch{2-})
 were done with HF and the BP86\cite{Becke1988,Perdew1986} and TPSSh XC functionals\cite{Tao2003}.
The scalar-relativistic zeroth-order regular approximation\cite{vanLenthe1993} (ZORA) 
 Hamiltonian was used together with def2-TZVPP basis set\cite{Weigend2005}
 with customized contraction coefficients for the ZORA Hamiltonian.\cite{PantazisZORA}
The geometry optimized in D\tief{2h} point-group symmetry with BP86/ZORA-TZVPP
 for the high-spin state ($M_S = 11$) and is provided as Supplementary Material.
The unrestricted broken-symmetry singlet calculations started from a unrestricted high-spin calculation
 with an subsequent exchange the alpha and beta density localized at one of the two iron atoms (\texttt{FlipSpin}).

The experimental crystal structure of the hemocyanin model complex
 was taken from Ref.\ \onlinecite{Kitajima1992}.
Only the position of the hydrogen atoms was optimized
 with an unrestricted triplet BP86\cite{Becke1988,Perdew1986} 
 DFT calculation using the def2-SVP basis set.\cite{Weigend2005,Weigend2006}
 and the D3BJ semi-empirical dispersion correction.\cite{Grimme2010}
The optimized structure is available as Supplementary Material.

Note that for reasons of convenience,
the maximum norm rather than the Frobenius norm is shown for DIIS and KDIIS in all figures
 because this norm is also used for those implementations.
However, TRAH works entirely with the Frobenius norm.

\section{Results and Discussion} \label{results}

\subsection{A benchmark on SCF convergence} \label{sec:scus-bench}

To demonstrate the robustness and efficiency of the new TRAH-SCF implementation
 for restricted and unrestricted SCF, we compare
 the number of iterations and the occurrence of erratic convergence behavior 
 with those of DIIS and KDIIS. 
For this purpose we chose a test set of small molecules
 that should be handled rather with multi-configurational methods
 and was initially proposed by Scuseria and his coworkers.\cite{Daniels2000,Kudin2002}

\begin{center}
\begin{figure}[t]
 \centering
 \includegraphics[width=.9\textwidth]{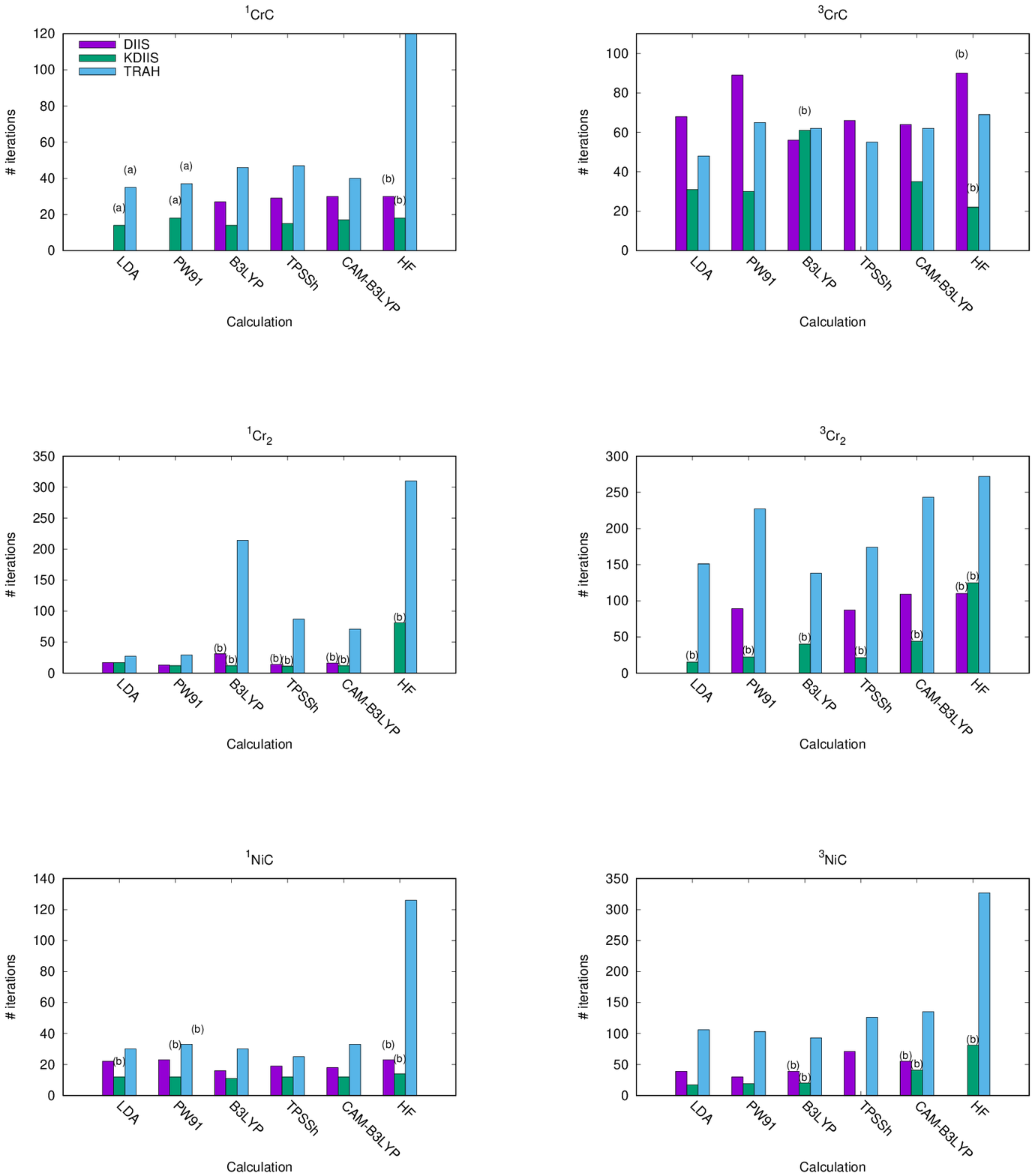}
 \caption{
   Number of SCF iterations using DIIS, KDIIS, and TRAH for singlet restricted (1) and
  triplet (3) unrestricted calculations on CrC, Cr\tief{2}, and NiC. 
   Diverging calculations are omitted;
   (a) and (b) denote convergence either to a doubly excited determinant
   or to a higher energy solution, respectively.
   For further details see text.
   }
 \label{fig:scuse-iter-1}
\end{figure}
\end{center}

\begin{center}
\begin{figure}[t]
 \centering
 \includegraphics[width=.9\textwidth]{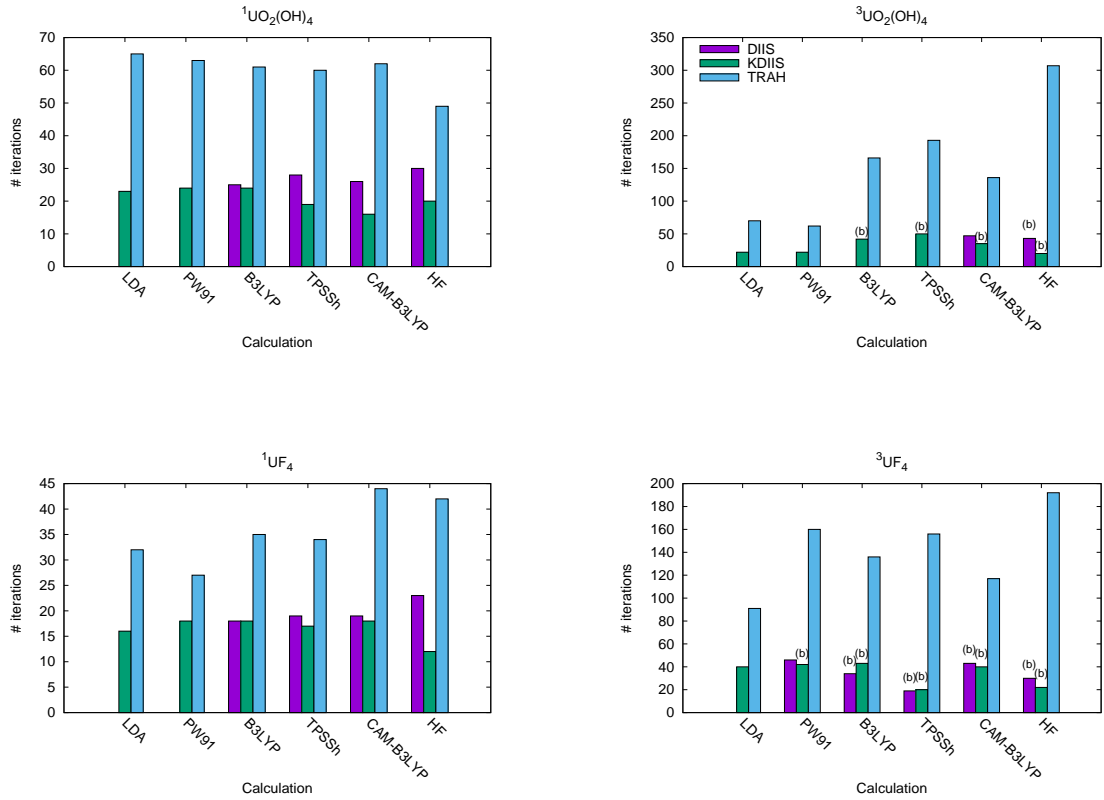}
 \caption{
   Number of SCF iterations using DIIS, KDIIS, and TRAH for singlet restricted (1) and 
  triplet unrestricted (3) calculations on UO\tief{2}(OH)\tief{4} and UF\tief{4}. 
   Diverging calculations are omitted;
   (b) denotes convergence to a higher energy solution.
   For further details see text.
 }
 \label{fig:scuse-iter-2}
\end{figure}
\end{center}

The number of SCF iterations needed to reach convergence with
 of DIIS, KDIIS, and TRAH using the default settings is given in Fig.\ \ref{fig:scuse-iter-1} and \ref{fig:scuse-iter-2}.
In general, we observe that KDIIS needs fewer iterations than DIIS which in turn needs less
 iterations than TRAH except of the ${^3}$CrC calculations.
For those calculations TRAH needs for most functionals less iterations to converge than DIIS.
The larger number of iterations of TRAH is expected because we have in every 
macro iteration several micro iterations that approximately diagonalize the augmented Hessian
until the residual norm is below $\gamma\, ||\bg||_2$.
Also we chose tighter convergence criteria for TRAH on purpose.

The erratic convergence behavior that we observed for the 60 SCF calculations
with each approach, i.e.\ DIIS, KDIIS, and TRAH, is summarized in Tab.\ \ref{tab:scuse-summary}.
Our new implementation of TRAH-SCF never diverged as expected
 and lead often to a solution with lower energy than with DIIS or KDIIS
 at the expense of significantly more TRAH iterations.
Only for a single calculation (${^1}$NiC with PW91) TRAH found a solution with a slightly higher energy than
 the DIIS solution.
Particularly worrisome in this regard are HF calculations.
Except for the closed-shell singlet calculation on UO\tief{2}(OH)\tief{2}, for which all variants converged
 to the same solution,
TRAH always found a solution with a lower energy than the competing DIIS and KDIIS methods.
All those unrestricted triplet solutions for which TRAH found a lower-energy solution
 were accompanied with a larger spin contamination as shown in Tab.\ \ref{tab:scuse-s2}.
For some calculations, e.g.\ ${^3}$UF\tief{4} with HF, it was barely noticeable.
However, for others, e.g.\ ${^3}$Cr\tief{2} with HF, the unrestricted 
$\langle S^2 \rangle$ expectation value of 6.23 indicated even 
 a potential quintet solution though a triplet calculation was performed.
Note that neither TRAH nor any DIIS algorithm can ensure 
 that the SCF solution with the lowest energy has been found because this is a global
 rather than a local optimization problem.

For the KDIIS and TRAH calculation on ${^1}$CrC with the LDA and PW91 functionals
 the energy minimization converged to an excited determinant solution
 for which the LUMO is doubly occupied and the HOMO is empty.
After swapping the HOMO and LUMO orbitals and restarting, convergence to the same
 excited-solution that violate the \textit{aufbau} principle was obtained.
Restarting the DIIS calculation from the TRAH solution with the swapped orbitals
 did not converge as before.
 
\subsection{Application to a transition-metal cluster}\label{sec:ru4co}

\begin{figure}
 \centering
 \includegraphics[width=.45\textwidth]{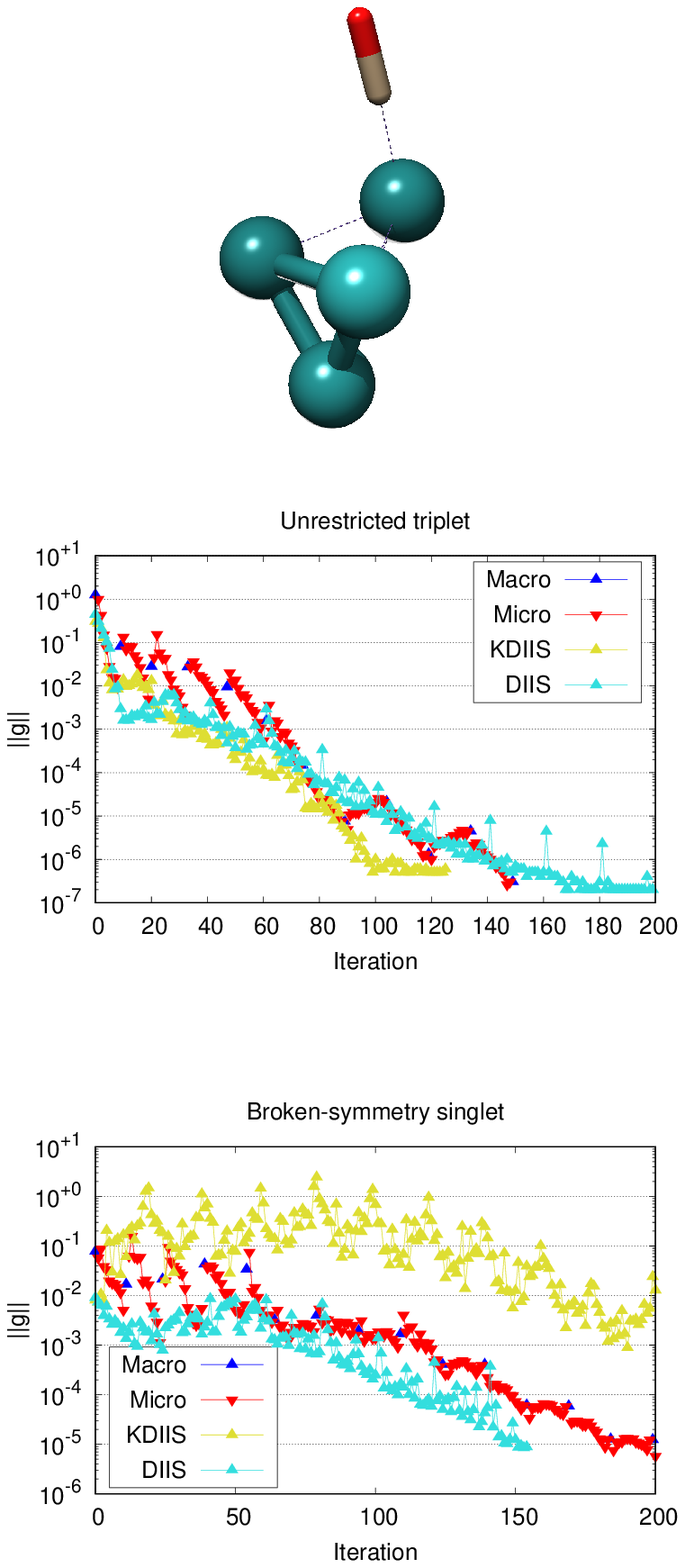}
 \caption{
   Number of SCF iterations using DIIS, KDIIS, and TRAH for unrestricted triplet 
     and broken-symmetry singlet calculations on Ru\tief{4}CO.
   For further details see text.
   }
 \label{fig:ru4co}
\end{figure}

The SCF convergence with various DIIS methods\cite{Hu2010,*Garza2012} was investigated previously
 for a small cluster model (Ru\tief{4}CO) of CO adsorption on Ru surfaces.\cite{Zeinalipour-Yazdi2008}
To relate our results to those of previous studies, we employed the B3LYP XC functional but investigated the SCF convergence
for both restricted and unrestricted wavefunctions.
Additionally, we also performed unrestricted calculations with higher multiplicities ($M_S=5,7,\ldots,13$).
In contrast to previous works,\cite{Hu2010,*Garza2012}
 none of our SCF calculations showed divergence.
That holds for the restricted calculations and all unrestricted calculations.
Triplet instabilities were detected for our restricted B3LYP/def2-TZVPP solutions.
In fact, we observed that the unrestricted calculation in a heptet spin state ($M_S=7$) 
 has the lowest energy; the unrestricted singlet and triplet ground state have a 0.42 and 0.27~eV higher energy, respectively.
All three SCF methods (DIIS, KDIIS, and TRAH) converged to the same 
 solution in case of the unrestricted calculations ($M_S=1,3,\ldots,13$).
The open-shell singlet B3LYP calculation showed the largest spin contamination (2.8)
and was the hardest to converge.
The gradient norm of DIIS-, KDIIS-, and TRAH-SCF open-shell singlet calculations
 is shown in Fig.\ \ref{fig:ru4co}.
Additionally, the convergence of the triplet calculations is presented in Fig.\ \ref{fig:ru4co}.

As shown in Fig.\ \ref{fig:ru4co}
 DIIS and TRAH show similar convergence rates if one also includes the
 micro iterations in the analysis.
The fewer number of SCF iterations in the DIIS calculations is mainly related
 to a faster convergence in the initial iterations rather convergence
 rates in a very shallow quadratic convergence region.
KDIIS showed the best convergence for triplet case but 
 showed a very slow convergence for the singlet calculation.
There the maximum gradient norm went below $10^{-3}$ only after 190 iteration.

\subsection{Application to a Fe\tief{2}S\tief{2} complex} \label{sec:roussin}

\begin{figure}
 \centering
 \includegraphics[width=.45\textwidth]{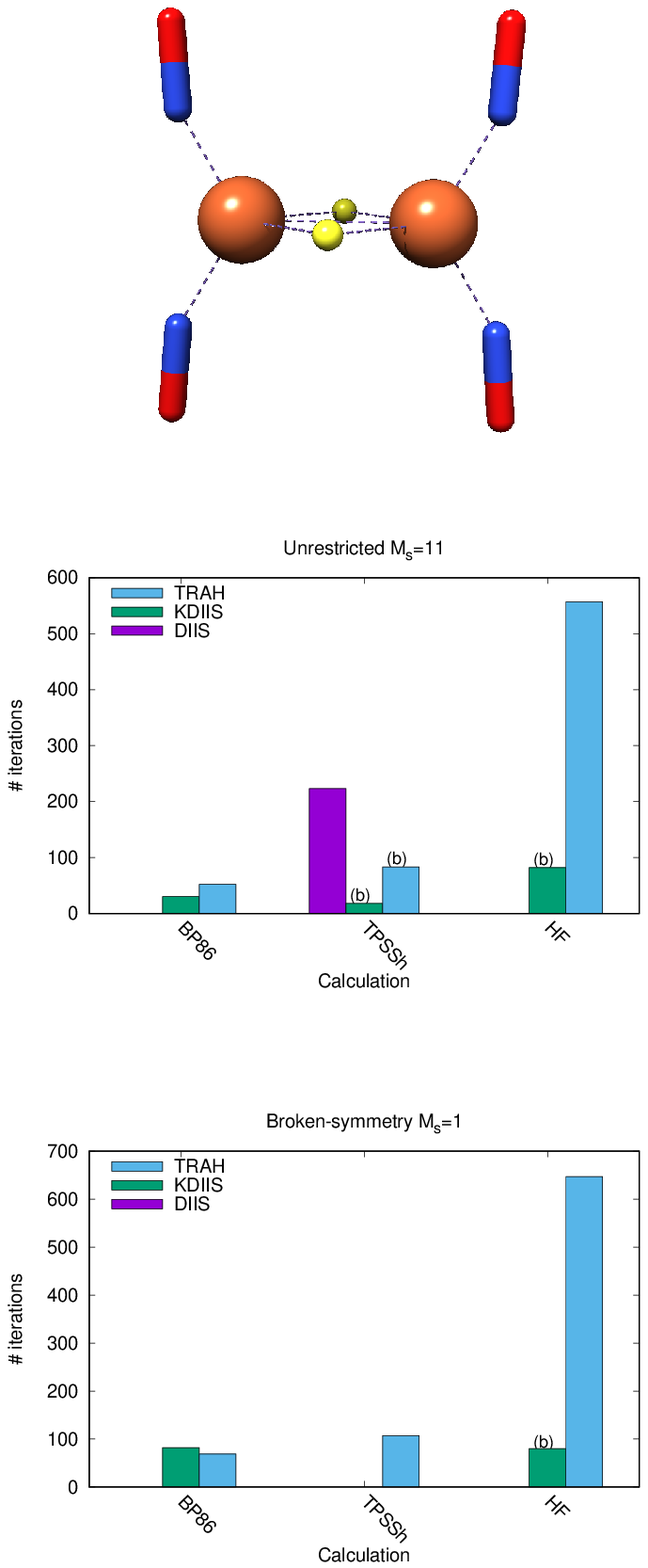}
 \caption{
   Number of SCF iterations using DIIS, KDIIS, and TRAH for unrestricted high-spin  ($M_S=11$)
     and broken-symmetry singlet calculations on Roussin's red dianion [Fe\tief{2}S\tief{2}(NO)\tief{4}]\hoch{2-}.
   (b) denotes convergence to a higher energy solution.
   For further details see text.
   }
 \label{fig:roussin}
\end{figure}

The SCF solution and the convergence behavior of DIIS, KDIIS, and TRAH
 is investigated for high-spin ($M_{S}=11$) and broken-symmetry singlet calculation
 of the Roussin's red salt dianion (see Fig.\ \ref{fig:roussin}).
The two iron centers are antiferromagnetically coupled as has been revealed in previous
 computational studies.\cite{Jaworska2005,*Hopmann2010}

Concerning our BP86 calculations, the high-spin DIIS calculation diverges
 and, hence, the subsequent broken-symmetry singlet calculation is not pursued.
KDIIS and TRAH do not show any converge issues for both the high-spin and singlet
 calculation and converge to the same energy.
According to Mulliken atomic charges, the point-group symmetry
  of both high-spin and BS low-spin solutions is intact.
With TRAH it takes (+22) more iterations than with KDIIS to converge the high-spin calculation.
For the singlet calculation (-13) less iteration are needed.

For the TPSSh high-spin calculations, all SCF implementations converged.
DIIS found a symmetry-broken solution that had a lower energy than
the one obtained with KDIIS and TRAH which is accompanied with
 much more iterations to reach convergence.
As observed already previously (Sec.\ \ref{sec:scus-bench}),
 the energy lowering is accompanied by larger spin contamination as can
 be seen from Tab.\ \ref{tab:roussin-s2}.
This broken-symmetry high-spin DIIS solution features an increased partial positive charge on
  one Fe atom and a reduced partial charge on the other.
The BS singlet calculation only converged with our new TRAH implementation.
For this BS singlet solution, the point-group symmetry is not broken
  as revealed by the Mulliken charges.
Since the hybrid meta-GGA TPSSh includes exact HF exchange,
 we observe a larger spin contamination in particular for the BS singlet calculation (see Tab.\ \ref{tab:roussin-s2}).

Finally, we have also performed HF calculations. 
Again, only KDIIS and TRAH converged with default settings.
For the high-spin calculation, TRAH breaks the point-group symmetry
 while converging to a solution with a lower energy than KDIIS.
Also for the broken-symmetry singlet calculation, TRAH converges to
 a solution with a lower energy than KDIIS and a larger deviation from the
 ideal $\langle S^2 \rangle$ value (see Tab.\ \ref{tab:roussin-s2}).
Both high-spin and singlet TRAH calculation take much mores iteration then KDIIS
 which is again attributed to the complicated energy landscape that
 offers the possibility to find many symmetry-broken solutions.

\subsection{Performance of TRAH for a hemocyanin model complex}\label{sec:hemocy}

\begin{figure}
 \centering
 \includegraphics[width=.45\textwidth]{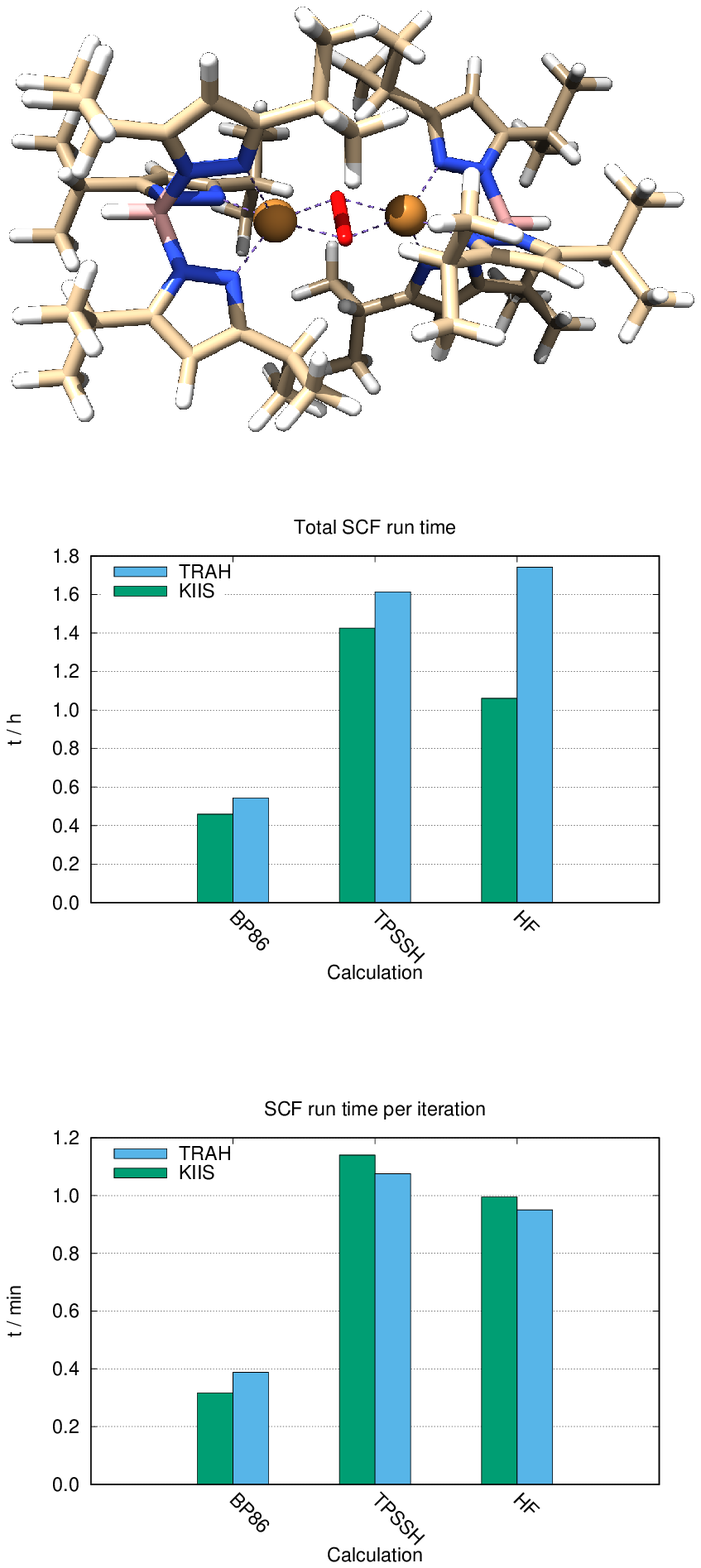}
 \caption{
   Total SCF timings and SCF timings per iteration for the KDIIS and TRAH for unrestricted triplet 
     and broken-symmetry singlet calculations on a hemocyanin model complex\cite{Kitajima1992}.
   For further details see text.
   }
 \label{fig:kitajima}
\end{figure}

After focusing on convergence characteristics
of our new TRAH-SCF implementation, we finally investigate the
 run-time performance in comparison with the other DIIS based convergers.
For this purpose, we calculated the unrestricted triplet and broken-symmetry
 singlet SCF energy of a hemocyanin model complex\cite{Kitajima1992}
 that has 164 atoms (see Fig.\ \ref{fig:kitajima}).
We employed for those calculations RI approximation for
 Coulomb matrix\cite{Baerends1973,*Dunlap1977,*Vahtras1993,*Neese2003} and the semi-numerical chain-of-spheres algorithm for
 exchange matrices\cite{Neese2009c,*Izsak2011,*Helmich-ParisUnpub}. 
As in the previous section, we used HF and the BP86 and TPSSh XC functionals.
Our SCF calculations with the def2-TZVPP orbital basis set\cite{Weigend2005}
 and def2/J auxiliary basis\cite{Weigend2006} set used 3586 and 4624 basis functions, respectively,
 and ran in parallel on a single Intel Haswell node (Intel{\textregistered} Xeon{\textregistered} CPU E5-2687W v3 @ 3.10~GHz)
 with 20 MPI processes.

We focus only on KDIIS and TRAH because the DIIS calculations
 did not converge for all SCF type calculations.
For all six calculations (unrestricted triplet and broken-symmetry singlet BP86, TPSSh, and HF)
  both KDIIS and TRAH converged to the same solution.
We observed convergence to an excited determinant (negative HOMO-LUMO gap of $\beta$ orbitals)
  for the BP86 triplet calculation.
Still, this calculation was considered for the performance analysis.
After swapping HOMO and LUMO,
 the restarted KDIIS and TRAH calculation converged to the same excited-determinant solution
 while DIIS diverged again.
The total SCF timings for the joint triplet and BS singlet
 calculation as well as the total timings per SCF iteration are presented in Fig.\ \ref{fig:kitajima}.
As expected, a TRAH-SCF calculation took always longer than the 
 corresponding KDIIS calculation which can be attributed to two reasons:
 (i) The time-determining step for SCF calculations is the
  integral-direct construction of Fock matrix\cite{Almloef1982} in the atomic-orbital (AO) basis
 (see Appendix).
 For DIIS schemes, Fock matrices are updated in every iteration from difference densities
  to improve the standard density-based Schwartz screening.\cite{Haeser1989}
 Such an implementation strategy is not available for the various Davidson-type micro
  iterations of TRAH for which the full non-incremental AO-Fock matrix must be built.
 (ii) The presence of nested iterations --- macro and micro --- in the TRAH algorithm
  usually leads to a larger number of total iterations than with DIIS type algorithms,
  even though, the TRAH macro iterations show quadratic convergence.
According to the results shown in Fig.\ \ref{fig:kitajima},
 the role of the more effective screening for KDIIS seems to
 be negligible because the timings per iteration hardly differ
 between KDIIS and TRAH.
The difference in the total SCF timings of DIIS and a second-order
 TRAH optimizations is primarily caused by the number of iteration to
 reach convergence.
This observation should not be generalized since integral screening
 is very dependent on the molecular geometry and size and also on
 the type of basis functions.

\section{Conclusions} \label{sec:concl}

In this work, we have presented a trust-region augmented Hessian (TRAH) implementation for restricted and 
 unrestricted Hartree--Fock and Kohn--Sham methods.
For several open-shell molecules for which it is notoriously complicated to converge
 the SCF equations, we have always reached convergence with tight
 convergence thresholds and a modest number of iterations when using TRAH-SCF.

In comparison with DIIS and KDIIS,
 TRAH usually took more iterations to converge because for every new set of orbitals
 the level-shifted Newton-Raphson equations are solved approximately and iteratively
 by means of an eigenvalue problem.
Nevertheless, the total runtime of TRAH-SCF is still competitive with the DIIS-based approaches
 and calculations on large molecules with extended basis sets are feasible without
 any additional technical restrictions.

Often we observed that TRAH-SCF finds a solution with a lower energy than
 DIIS and KDIIS. 
For unrestricted calculations, this is accompanied by a larger spin contamination, i.e.\ larger
 deviation from the desired spin-restricted $\langle S^2 \rangle$ expectation value.
However, we found also rare cases in which DIIS found a solution with a lower energy than KDIIS and TRAH,
 which had one or more types of broken symmetries.
Furthermore, we observed that TRAH-SCF as well as KDIIS may also converge
 to an excited state determinant solution with a negative HOMO-LUMO gap.
For those calculations DIIS always diverged.
Though these solutions are not desired, they are valid energy minima 
 with respect to variations of the orbital coefficients.

Future research projects will be devoted to finding ways for
 a restrained optimization for broken-symmetry determinants\cite{Herrmann2009}
 by using a modified Lagrangian with the TRAH algorithm.
Likewise convergence to excited-state determinants could be prohibited.
An extension to restricted open-shell and MC SCF theories
 will be pursued as well in our group.

\section{Supplementary Material}
See supplementary material for Cartesian coordinates of UO\tief{2}(OH)\tief{4},
 Roussin's red dianion [Fe\tief{2}S\tief{2}(NO)\tief{4}]\hoch{2-}, and the hemocyanin
 model complex.

\section{Acknowledgments}
The author acknowledges gratefully financial support from 
the Max Planck Society

\section{Appendix}

The TRAH-SCF implementation necessitates computations of electronic
gradients and linear transformations of the electronic Hessians (sigma vectors).
Since most calculations presented in this work were open-shell calculations,
 explicit expressions for the electronic gradient
\begin{align}
 g_{ai}^{\sigma} &= \bra{0} [ X^{-,\sigma}_{ai}, \hat{H} ] \ket{0} = - F^{\sigma}_{ai}  \\
 &= - \sum_{\mu\nu} C_{\mu a}^{\sigma} \, F^{\sigma}_{\mu\nu}[\bD^{\alpha},\bD^{\beta}] \, C^{\sigma}_{\nu_i} \\
X^{-,\sigma}_{ai} &=  a^{\dag}_{a\sigma}a_{i\sigma} - a^{\dag}_{i\sigma}a_{a\sigma}
\end{align}
and the sigma vector
\begin{align}
\sigma_{ai}^{\sigma} &= \sum_{\tau} \sum_{bj} 
 \frac{1}{2}\bra{0} [ X^{-,\sigma}_{ai},  [ X^{-,\tau}_{bj}, \hat{H} ] ] \notag \\
&\phantom{= \sum_{\tau} \sum_{bj} \frac{1}{2}\bra{0}}
                   + [ X^{-,\tau}_{bj},  [ X^{-,\sigma}_{ai}, \hat{H} ] ] \ket{0} \, \kappa^{\tau}_{bj} \\
 &= - \sum_{\mu\nu} ( 
  C_{\mu a}^{\sigma} \, F_{\mu\nu}[\bD^{\alpha},\bD^{\beta}] \, \overline{\Lambda}^{h,{\sigma}}_{\nu i} \notag \\
  & \phantom{- \sum_{\mu\nu}}    +\overline{\Lambda}^{p,\sigma}_{\mu a} \, F_{\mu\nu}[\bD^{\alpha},\bD^{\beta}] \, C^{\sigma}_{\nu i} ) \notag \\
 & \phantom{- \sum_{\mu\nu}} + C_{\mu a}^{\sigma} \overline{F}_{\mu\nu}^{\sigma}[\overline{\bD}^{\alpha},\overline{\bD}^{\beta}] C^{\sigma}_{\nu i} )
\end{align}
are given for spin-unrestricted SCF wavefunctions.
We follow the convention that the indices $i,j,\ldots$ denote occupied orbitals and $a,b,\ldots$ virtual orbitals.
Both equations are formulated in terms of Fock and Kohn-Sham potential matrices in the sparse
 AO basis represented by Greek-letter indeces.
The following intermediates are needed to compute the gradient
and sigma vectors with an AO-based formulation:
\begin{align}
&F_{\mu\nu}^{\sigma}[\bD^{\alpha},\bD^{\beta}] 
 = h_{\mu\nu} + G_{\mu\nu}^{\sigma}[\bD^{\alpha},\bD^{\beta}] \notag \\
 & \phantom{F_{\mu\nu}^{\sigma}[\bD^{\alpha},\bD^{\beta}] = h_{\mu\nu}}
  + c^{xc} \, V^{\sigma,\xc}_{\mu\nu}[\bD^{\alpha},\bD^{\beta}] \\
& G_{\mu\nu}^{\sigma}[\bD^{\alpha},\bD^{\beta}] =  \sum_{\kappa\lambda} \Big( 
  D_{\kappa\lambda} (\mu\nu|\kappa\lambda)  \notag \\
 &\phantom{G_{\mu\nu}^{\sigma}[\bD^{\alpha},\bD^{\beta}] =  \sum_{\kappa\lambda} \Big(}
  - c^X \, D^{\sigma}_{\kappa\lambda} \, (\mu\lambda|\kappa\nu) \Big) \\
&V^{\xc,\sigma}_{\mu\nu}[\bD^{\sigma}] \overset{\text{LDA}}{=} \int d\br \left[ \frac{\delta e^{\xc}}{\delta \rho^{\sigma}(\br)} \chi_{\mu}(\br) \chi_{\nu}(\br) \right] \\
&\overline{F}_{\mu\nu}^{\sigma}[\bD^{\alpha},\bD^{\beta}] 
 = G_{\mu\nu}^{\sigma}[\bD^{\alpha},\bD^{\beta}] \notag \\
 & \phantom{F_{\mu\nu}^{\sigma}[\bD^{\alpha},\bD^{\beta}] = }
  + c^{xc} \, \overline{V}^{\sigma,\xc}_{\mu\nu}[\bD^{\alpha},\bD^{\beta}] \\
&\overline{V}^{\xc,\sigma}_{\mu\nu}[\bD^{\alpha},\bD^{\beta}] \overset{\text{LDA}}{=} \notag \\
&\int d\br \, d\br' \sum_{\tau} \left[ \frac{\delta^2 e^{\xc}}{\delta \rho^{\sigma}(\br)\, \delta \rho^{\tau}(\br')} \chi_{\mu}(\br) \chi_{\nu}(\br) \, \rho(\br',\bD^{\tau}) \right]
\end{align}
For the intermediates above the following densities $\rho$, 
 density matrices $\bD$, and trial vector-containing MO coefficients\cite{Koch1996,*Haettig2000} $\overline{\boldsymbol{\Lambda}}$
 are used:
\begin{align}
&\rho(\br,\bD^{\tau}) = \sum_{\mu\nu} \chi_{\mu}(\br) \, D^{\tau}_{\mu\nu} \, \chi_{\nu}(\br) \\
& D^{\sigma}_{\mu\nu} = \sum_k C_{\mu k}^{\sigma}\, C_{\nu k}^{\sigma} \\
& D_{\mu\nu} = D^{\alpha}_{\mu\nu} +  D^{\beta}_{\mu\nu} \\
& \overline{D}^{\sigma}_{\mu\nu} = \sum_k ( \overline{\Lambda}_{\mu k}^{h,\sigma}\, C_{\nu k}^{\sigma}
 +  C_{\mu k}^{\sigma}\, \overline{\Lambda}_{\nu k}^{h,\sigma} ) \\
& \overline{\Lambda}_{\mu a}^{p,\sigma} =  \phantom{-} \sum_kC_{\mu k}^{\sigma} \, \kappa^{\sigma}_{ak} \\
& \overline{\Lambda}_{\mu i}^{h,\sigma} = -\sum_c C_{\mu c}^{\sigma} \, \kappa^{\sigma}_{ci}
\end{align}
For simplicity reasons, the XC potential matrices are only given for a local density
 approximation (LDA). All other functionals that are available in ORCA
 can be employed as well with TRAH-SCF.
Though not discussed in this work, please note that solvation effects
 can also be accounted for in the new TRAH-SCF implementation 
 with the conductor-like polarizable continuum model\cite{Tomasi2005,*Barone1998} (C-PCM).

\bibliography{references}


\begin{table}
 \caption{Default parameters of the TRAH-SCF implementation.}
 \label{tab:para}
 \begin{ruledtabular}
 \begin{tabular}{lr}
 Electronic gradient norm for converging & $10^{-6}$ \\
macro iterations & \\
 Maximum number of micro iterations      &    16 \\ 
 Maximum number of macro iterations   &    64 \\
 Number of Davidson start vectors     &     2 \\
 Gradient scaling factor $\gamma$ for micro iteration&   0.1 \\
 accuracy  & \\
 Minimum micro iteration accuracy    &   0.01 \\
 Gradient norm threshold for switching to NR   &   $10^{-3}$ \\
 Initial trust radius                 &   0.4 \\
 Minimum AH scaling parameter ($\alpha_{\text{min}}$)    &   1 \\
 Maximum AH scaling parameter ($\alpha_{\text{max}}$)    & 1000
 \end{tabular}
 \end{ruledtabular}
\end{table}

\begin{table}
 \caption{ Summary of erratic convergence behavior.}
 \label{tab:scuse-summary}
 \begin{ruledtabular}
 \begin{tabular}{lrrr}
  & 
 \multicolumn{1}{c}{DIIS} & 
 \multicolumn{1}{c}{KDIIS} & 
 \multicolumn{1}{c}{TRAH} \\
Divergence & 15 & 2 & 0 \\
Doubly excited determinant & 0 & 2 & 2 \\
Higher-energy state & 14 &  19 & 1
 \end{tabular}
 \end{ruledtabular}
\end{table}

\begin{table}
 \caption{ $\langle S^{2} \rangle$ UKS / UHF expectation values with DIIS, KDIIS, and TRAH SCF for $M_s = 3$. 
    No convergence is denoted by /.}
 \label{tab:scuse-s2}
 \begin{ruledtabular}
 \begin{tabular}{lrrr}
 \multicolumn{1}{c}{Mol.} & 
 \multicolumn{1}{c}{DIIS} & 
 \multicolumn{1}{c}{KDIIS} & 
 \multicolumn{1}{c}{TRAH} \\
 & \multicolumn{3}{c}{LDA} \\[0.5em]
$^{3}$CrC            & 3.94 & 3.94 & 3.94 \\
$^{3}$Cr$_{2}$       & /    & 2.00\footnotemark[2] & 3.75 \\
$^{3}$NiC            & 2.03 & 2.03 & 2.03 \\
$^{3}$UO$_2$(OH)$_4$ & /    & 2.01 & 2.01 \\
$^{3}$UF$_{4}$       & /    & 2.00 & 2.00 \\[1.0em]
 & \multicolumn{3}{c}{PW91} \\[0.5em]
$^{3}$CrC            & 4.14 & 4.15 & 4.14 \\
$^{3}$Cr$_{2}$       & 4.07 & 2.01\footnotemark[2] & 4.07 \\
$^{3}$NiC            & 2.02 & 2.02 & 2.02 \\
$^{3}$UO$_2$(OH)$_4$ & /    & 2.01 & 2.01 \\
$^{3}$UF$_{4}$       & 2.00 & 2.00\footnotemark[2] & 2.00 \\[1.0em]
 & \multicolumn{3}{c}{B3LYP} \\[0.5em]
$^{3}$CrC            & 4.42 & 3.34\footnotemark[2] & 4.42 \\
$^{3}$Cr$_{2}$       & /    & 2.02\footnotemark[2] & 4.90 \\
$^{3}$NiC            & 2.09\footnotemark[2] & 2.01\footnotemark[2] & 2.52 \\
$^{3}$UO$_2$(OH)$_4$ & /    & 2.03\footnotemark[2] & 2.05 \\
$^{3}$UF$_{4}$       & 2.00\footnotemark[2] & 2.00\footnotemark[2] & 2.00 \\[1.0em]
 & \multicolumn{3}{c}{TPSSh} \\[0.5em]
$^{3}$CrC            & 4.44 & /    & 4.45 \\
$^{3}$Cr$_{2}$       & 4.77 & 2.01\footnotemark[2] & 4.77 \\
$^{3}$NiC            & 2.42 & /    & 2.42 \\
$^{3}$UO$_2$(OH)$_4$ & /    & 2.03\footnotemark[2] & 2.04 \\
$^{3}$UF$_{4}$       & 2.00\footnotemark[2] & 2.00\footnotemark[2] & 2.00 \\[1.0em]
 & \multicolumn{3}{c}{CAM-B3LYP} \\[0.5em]
$^{3}$CrC            & 4.37 & 3.21\footnotemark[2] & 4.37 \\
$^{3}$Cr$_{2}$       & 4.73 & 2.02\footnotemark[2] & 4.73 \\
$^{3}$NiC            & 2.14\footnotemark[2] & 2.01\footnotemark[2] & 2.46 \\
$^{3}$UO$_2$(OH)$_4$ & 2.04 & 2.04\footnotemark[2] & 2.04 \\
$^{3}$UF$_{4}$       & 2.00\footnotemark[2] & 2.00\footnotemark[2] & 2.00 \\[1.0em]
 & \multicolumn{3}{c}{HF} \\[0.5em]
$^{3}$CrC            & 4.77\footnotemark[2] & 2.05\footnotemark[2] & 4.88 \\
$^{3}$Cr$_{2}$       & 5.37\footnotemark[2] & /    & 6.23 \\
$^{3}$NiC            & /    & 2.33\footnotemark[2] & 3.73 \\
$^{3}$UO$_2$(OH)$_4$ & 2.14\footnotemark[2] & 2.14\footnotemark[2] & 2.02 \\
$^{3}$UF$_{4}$       & 2.01\footnotemark[2] & 2.01\footnotemark[2] & 2.01
 \end{tabular}
 \end{ruledtabular}
\footnotetext[2]{Convergence to higher-energy state.}
\end{table}

\begin{table}
 \caption{ $\langle S^{2} \rangle$ UKS / UHF expectation values 
 for Roussin's red dianion with DIIS, KDIIS, and TRAH SCF. 
    No convergence is denoted by /.}
 \label{tab:roussin-s2}
 \begin{ruledtabular}
 \begin{tabular}{lrrr}
 \multicolumn{1}{c}{$M_S$} & 
 \multicolumn{1}{c}{DIIS} & 
 \multicolumn{1}{c}{KDIIS} & 
 \multicolumn{1}{c}{TRAH} \\
 & \multicolumn{3}{c}{BP86} \\[0.5em]
11 & / & 30.06 & 30.06 \\
 1 & / &  2.82 &  2.82 \\[1.0em]
 & \multicolumn{3}{c}{TPSSh} \\[0.5em]
11 & 30.33 & 30.19\footnotemark[2] & 30.19\footnotemark[2] \\
 1 & /    & / & 4.53 \\[1.0em]
 & \multicolumn{3}{c}{HF} \\[0.5em]
11 & / & 32.52\footnotemark[2] & 33.70 \\
 1 & / &  7.31\footnotemark[2] &  8.36 \\[1.0em]
 \end{tabular}
 \end{ruledtabular}
\footnotetext[2]{Convergence to higher-energy state.}
\end{table}

\end{document}